# THE PHQMD MODEL FOR THE FORMATION OF NUCLEAR CLUSTERS AND HYPERNUCLEI IN HEAVY-ION COLLISIONS

V. Kireyeu[1,*], J. Aichelin[2], E. Bratkovskaya[3,4], A. Le Fèvre[3], V. Lenivenko[1],

V. Kolesnikov[1], Y. Leifels[3], V. Voronyuk[1]

[1,*]*Joint Institute for Nuclear Research, Dubna, Russia*

[2]*Université de Nantes, France*

[3]*GSI Helmholtzzentrum für Schwerionenforschung GmbH, Darmstadt, Germany*

[4]*Institut fur Theoretische Physik, Johann Wolfgang Goethe-Universitat*

*E-mail: vkireyeu@jinr.ru

**Abstract** — Modeling of the process of the formation of nuclear clusters in the hot nuclear matter is a challenging task. We present the novel n-body dynamical transport approach - PHQMD (Parton-Hadron-Quantum-Molecular Dynamics) [1] for the description of heavy-ion collisions as well as clusters and hpernuclei formation. The PHQMD extends well established PHSD (Parton-Hadron-String Dynamics) approach - which incorporates explicit partonic degrees-of-freedom (quarks and gluons), an equation-of-state from lattice QCD, as well as dynamical hadronization and hadronic elastic and inelastic collisions in the final reaction phase, by n-body quantum molecular dynamic propagation of hadrons which allows choosing of the equation of state with different compression modulus. The formation of clusters, including hypernuclei, is realized by incorporation the Simulated Annealing Clusterization Algorithm (SACA). We present first results from PHQMD on the study of the production rates of strange hadrons, nuclear clusters and hypernuclei in elementary and heavy-ion collisions at NICA energies. In particular, sensitivity on the "hard" and "soft" equation of state within the PHQMD model was investigated for "bulk" observables.

INTRODUCTION

New experiments NICA-MPD and BM@N in Dubna and CBM at FAIR are under construction with the aim to investigate the QCD phase diagram at low temperatures and high baryon densities by studying heavy-ion collisions in the energy range $\sqrt{s_{NN}} < 11$ GeV.

Heavy-ion collisions provide a unique possibility to create and investigate hot and dense matter in the laboratory. At the initial stage of the reaction at relativistic incident energy a new state of matter, a quark-gluon plasma (QGP) is formed, while the final stage is driven by the hadronization process and the formation of clusters. The capture of the produced hyperons by clusters of nucleons leads to the hypernuclei formation which is a very rare process when the reaction occurs at strangeness threshold energies. The dynamical formation of fragments may lead to a more accurate description of observables like flow harmonics or transverse momentum spectra. It may also help to explore the new physics opportunities like hypernuclei formation, 1-st order phase transition, fragment formation at (ultra-) relativistic energies.

THE PHQMD MODEL

The PHQMD is a novel n-body transport approach [1] which unites the collision integrals of the Parton-Hadron-String Dynamics (PHSD) [2] approach with 2-body potential interactions between baryons similar as in the Quantum Molecular Dynamics (QMD) [3] approach where baryons are described by Gaussian wave functions. While high energy inelastic hadron-hadron collisions in PHQMD (as in the PHSD) are described by the FRITIOF 7.02 and PYTHIA (v. 6.4) string models, the transport approach at low energy hadron-hadron collisions is based on experimental cross sections and matched to reproduce the nucleon-nucleon, meson-nucleon and meson-meson cross section data in a wide kinematic range. In the mean-field mode the initialization in coordinate space is realized by the point-like test particles, randomly redistributed according to the Wood-Saxon density distribution while in momentum space – according to the Thomas-Fermi distribution in the rest frame of the nucleus. In the QMD mode the single-particle Wigner density is used. The interaction between the nucleons has two parts, a local Skyrme type interaction and a Coulomb interaction:

$$V_{i,j} = V\left(\vec{r}_i, \vec{r}_j, \vec{r}_{i0}, \vec{r}_{j0}, t\right) = V_{\text{Skyrme}} + V_{\text{Coul}}$$
$$= \frac{1}{2}t_1 \delta\left(\vec{r}_i - \vec{r}_j\right) + \frac{1}{\gamma+1}t_2 \delta\left(\vec{r}_i - \vec{r}_j\right) \rho^{\gamma-1}\left(\vec{r}_i, \vec{r}_j, \vec{r}_{i0}, \vec{r}_{j0}, t\right)$$
$$+ \frac{1}{2}\frac{Z_i Z_j e^2}{\left|\vec{r}_i - \vec{r}_j\right|}. \quad (1)$$

For the Skyrme potential the analytical form is used:

$$\langle V_{\text{Skyrme}}(\vec{r}_{i0},t)\rangle = \alpha\left(\frac{\rho_{\text{int}}(\vec{r}_{i0},t)}{\rho_0}\right) + \beta\left(\frac{\rho_{\text{int}}(\vec{r}_{i0},t)}{\rho_0}\right). \quad (2)$$

Where $\rho_{\text{int}}$ is the interaction density obtained by convoluting the density distribution of particle with the distribution functions of all other particles of the surrounding medium.

For a given value of $\gamma$ the parameters $t_1$, $t_2$ in Equation 1 are uniquely related to the coefficients $\alpha$, $\beta$ of the Equation 2. Parameter sets for the nuclear equation of state used in the PHQMD model can be found in the Table.

An equation-of-state with a rather low value of the compression modulus K yields a weak repulsion against the compression of nuclear matter and thus describes "soft" matter (denoted by "S"). A high value of K causes a strong repulsion of nuclear matter under compression (called a hard EoS, "H").

## CLUSTERS FORMATION

The PHQMD approach conserves the correlations in the system and does not suppress fluctuations. Since clusters are *n*-body correlations, this approach is well suited to study the creation of clusters and its time evolution.

The simplest way to identify clusters is to use coalescence or minimum spanning tree procedures. The first needs a multitude of free parameters, the second may be used only for the identification at the end of the reaction when groups of nucleons are well separated in the coordinate space, and this excludes any study on the physical origin of the cluster formation [4].

To get over the limitation that fragments can be identified only at the final stage of the reaction it is possible to use the coordinate space information together with the momentum. This idea was launched by Dorso et al. [5] and it has been developed into the Simulated Annealing Clusterization Algorithm (SACA) [6].

The SACA algorithm consists of the following steps: at first the algorithm takes the positions and momenta of all nucleons at time t to determine clusters within a phase space coalescence approach using the Minimum Spanning Tree technique (MST). In the second step, the MST clusters and individual particles are recombined in all possible ways into fragments or they are left as single nucleons, such as to choose that configuration which has the highest binding energy. This procedure is repeated very many times within the Metropolis procedure and it automatically leads to the most bound configuration. We note that in the spectator region of the collision clusters chosen that way at early times are the pre-fragments of the final state clusters, because fragments are not a random collection of nucleons at the end but initial-final state correlations.

## MODEL RESULTS

*Elementary reactions*

In order to be conclusive on the particle production in heavy-ion collisions, the production of hadrons in elementary reactions has to be under control. That is not a trivial task due to the lack of experimental information, especially on multi-strange hyperons in the NICA energy range. Moreover, because in a heavy-ion interaction both protons and neutrons take part in the collision, it is very important to have the correct isospin decomposition for the hadrons production in *p+p*, *p+n* and *n+n* reactions. As an example, Fig. 1 show the energy dependence of the mean multiplicity of positively charged pions (Fig. 1*a*) and kaons (Fig. 1*b*) from in *p+p*, *p+n* and *n+n* inelastic collisions. Black circles represents compilation of the world experimental data for *p+p* collisions [7-14]. The PHQMD model predictions are drawn with solid lines. Since for elementary interactions the PHQMD contains a number of modifications - "tunes", adopted from the PHSD, of the underlying string model (FRITIOF 7.02 and PYTHIA6.4), the results from the default PYTHIA 8.2 (dashed lines) are included for reference, too. As seen from Fig. 1, the PHQMD with the "PHSD tune" describes reasonably well the experimental data on pion and kaon multiplicities in *p+p* collisions.

*Heavy-ion collisions*

The transverse mass spectra of protons, anti-protons and produced mesons in comparison with the STAR experimental data from Au+Au collisions at $\sqrt{s_{NN}}=11.5$ GeV [15] for "hard" and "soft" EoS are shown on Fig. 2-3 respectively. The centrality dependence of the spectra of newly produced particles is well described, a hard EoS increases the slope of the spectra at large $p_T$ as compared to a soft EoS, while the proton slope is still slightly underestimated at large $p_T$.

*Hypernuclei formation*

Fig. 4 shows rapidity distributions of $Z = 1$, $Z = 2$ charged particles, heavier clusters ($Z > 2$), $\Lambda$ hyperons, light ($A \leq 4$) and heavy ($A > 4$) hypernuclei identified by the MST algorithm for Au+Au collisions at 10 A·GeV. We observe enhanced yields of heavier fragments and hypernuclei close to the target and projectile rapidity regions, and almost a constant distribution of $Z = 1$ particles in between. At mid-rapidity only a small amount of hyperons are bound in small hypernuclei, unlike the projectile/target rapidities where many of produced hyperons are bound in larger hypernuclei.

CONCLUSIONS

We present the basic ideas of the PHQMD – an *n*-body transport approach, presently under active development, which combines mean-field and QMD propagation of baryons in one code allowing to study different descriptions of heavy-ion collisions. The collision integrals are adopted from the PHSD approach while the cluster recognition algorithm is realized by SACA model or the simpler MST technique. We demonstrate that the PHQMD within the "PHSD tune" of string model shows a good agreement with experimental data on pion and kaon multiplicities for elementary pp reactions in the NICA energy range. We find that the transverse momenta spectra of hadrons in heavy-ion collisions at NICA energies are sensitive to the equation of state of nuclear matter(EoS). The PHQMD model recognize clusters at mid-rapidity, as well as in the target/projectile region.

Predictions for clusters and hypernuclei formation for future NICA experiments are currently being made.

This work was supported by the Russian Science Foundation grant 19-42-04101 and the Deutsche Forschungsgemeinschaft (DFG, German Research Foundation).

FIGURE CAPTIONS

For the manuscript

V. Kireyeu, J. Aichelin, E. Bratkovskaya, A. Le Fèvre, V. Lenivenko,

V. Kolesnikov, Y. Leifels, V. Voronyuk

THE PHQMD MODEL FOR THE FORMATION OF NUCLEAR

CLUSTERS AND HYPERNUCLEI IN HEAVY-ION COLLISIONS

**Fig. 1.** Mean multiplicity of $\pi^+$ (panel *a*) and $K^+$ (panel *b*) produced in inelastic collisions. PHSD model predictions were drawn with red solid ("*p + p*"), blue dotted ("*p + n*") and green dot dashed ("*n + n*") lines, PYTHIA - with red dashed ("*p + p*"), blue dot dot dashed ("*p + n*") and green dashed ("*n + n*") lines. Black circles represent a compilation of the world experimental data for "*p + p*" collisions [7-14].

**Fig. 2.** Transverse momentum spectra at mid-rapidity of $\pi^+$ (*a*), $\pi^-$ (*b*), $K^+$ (*c*), $K^-$ (*d*), $p$ (*e*) and $\bar{p}$ (*f*) for Au + Au at GeV from PHQMD with "hard" EoS in comparison to the STAR experimental data (stars) from Ref. [15] for different centrality classes. The spectra for different centralities are multiplied by corresponding factors for better visibility: 0-5%·1 (circles); 5-10%·$10^{-1}$ (squares); 10-20%·$10^{-2}$ (triangles); 20-30%·$10^{-3}$ (diamonds); 30-40%·$10^{-4}$ (crosses); 40-50%·$10^{-5}$ (rotated crosses); 50-60%·$10^{-6}$ (rotated crossed squares).

**Fig. 3.** Same as Fig 2. with a "soft" EoS.

**Fig. 4.** PHQMD predictions (with a hard EoS and the MST algorithm) for the rapidity distributions of all charges (red line), $Z = 1$ particles (red dashed line), $Z = 2$

clusters (green dotted line), $Z > 2$ (orange dot-dot-dashed line), $\Lambda$'s (magenta line with triangles) as well as light hypernuclei with $A \leq 4$ (blue line with stars) and heavy hypernuclei with $A > 4$ (green line with dots) as a function of the rapidity for central Au+Au collisions at 10 A·GeV.

TABLE

For the manuscript

V. Kireyeu, J. Aichelin, E. Bratkovskaya, A. Le Fèvre, V. Lenivenko,

V. Kolesnikov, Y. Leifels, V. Voronyuk

THE PHQMD MODEL FOR THE FORMATION OF NUCLEAR

CLUSTERS AND HYPERNUCLEI IN HEAVY-ION COLLISIONS

**Table.** Parameter sets for the nuclear equation of state used in the PHQMD model.

|   | $\alpha$ (MeV) | $\beta$ (MeV) | $\gamma$ | K (MeV) |
|---|---|---|---|---|
| S | −390 | 320 | 1.14 | 200 |
| H | −130 | 59 | 2.09 | 380 |

Fig. 1 for manuscript „THE PHQMD MODEL FOR THE FORMATION OF NUCLEAR CLUSTERS AND HYPERNUCLEI IN HEAVY-ION COLLISIONS"

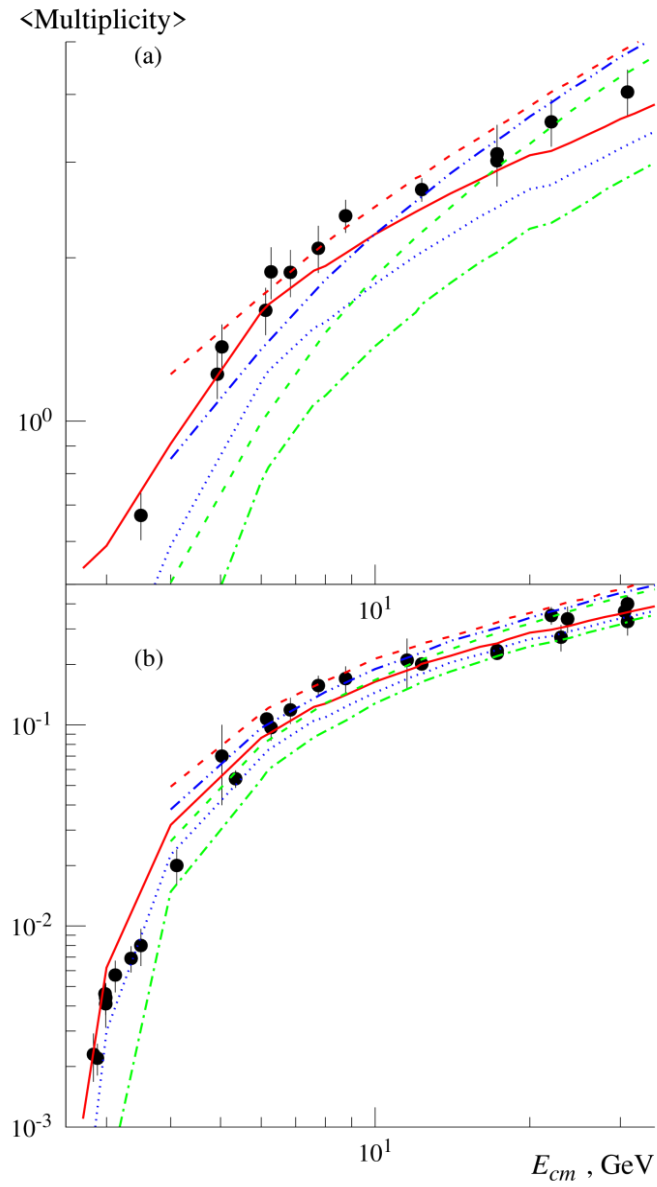

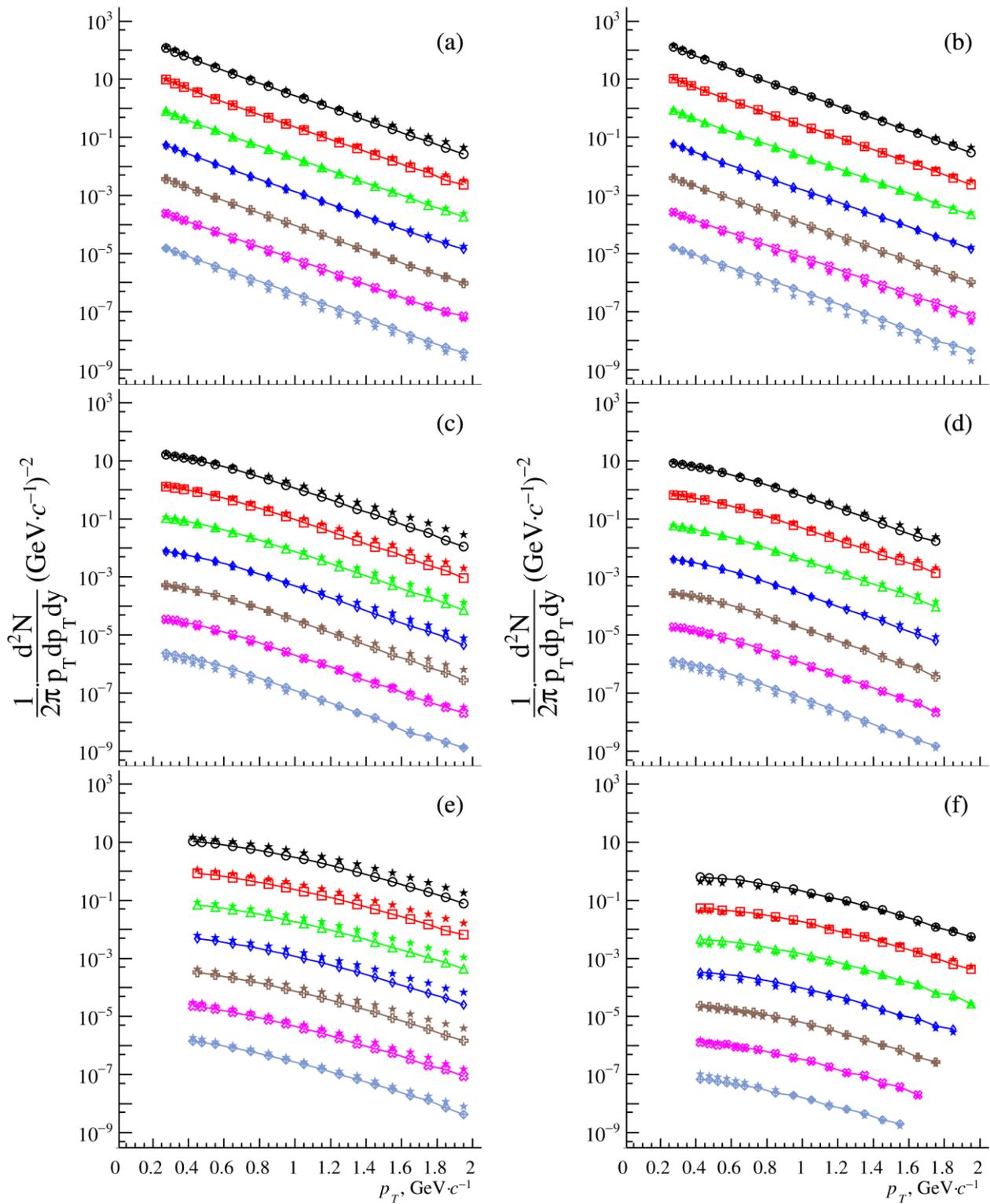

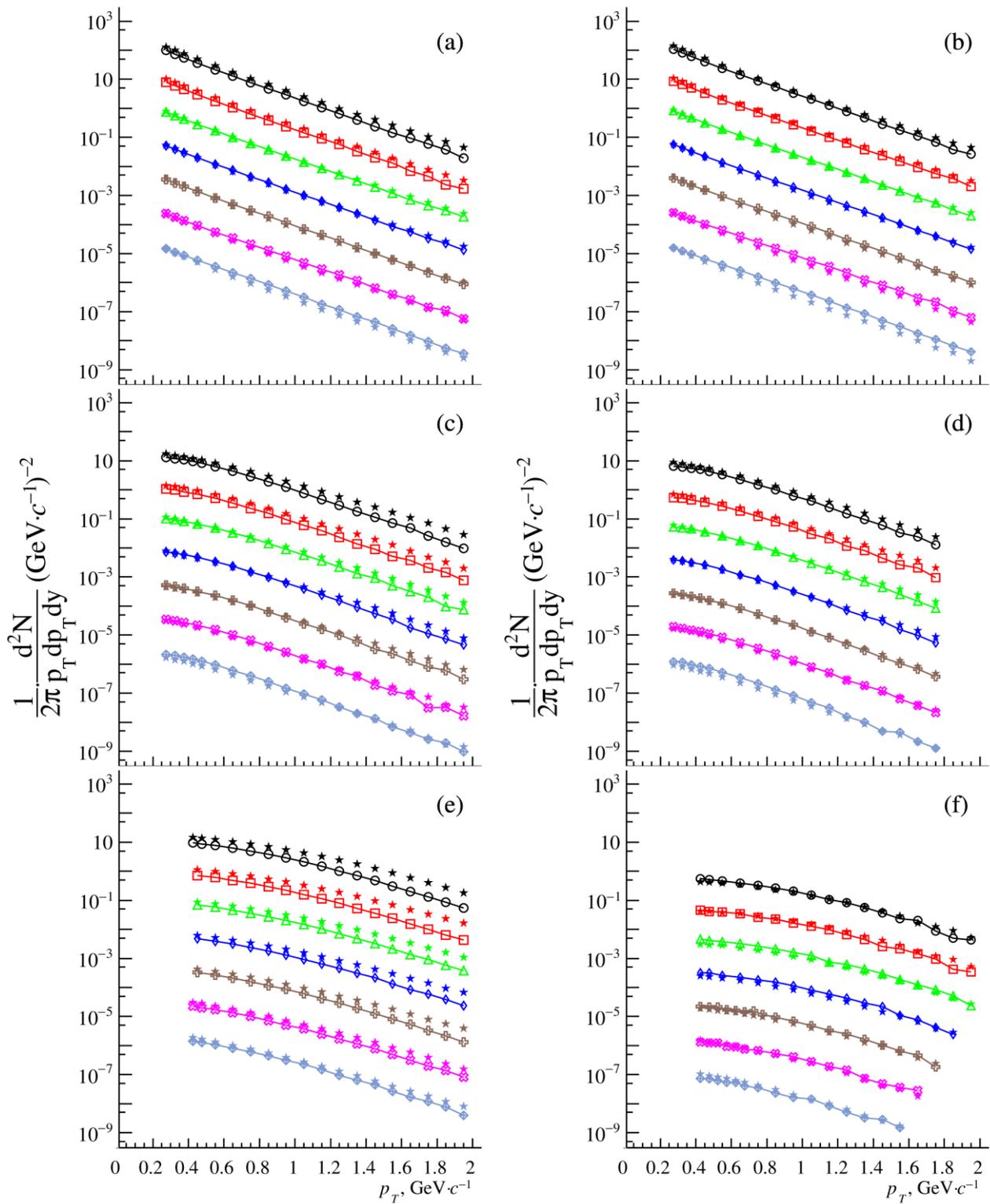

Fig. 4 for manuscript „THE PHQMD MODEL FOR THE FORMATION OF NUCLEAR CLUSTERS AND HYPERNUCLEI IN HEAVY-ION COLLISIONS"

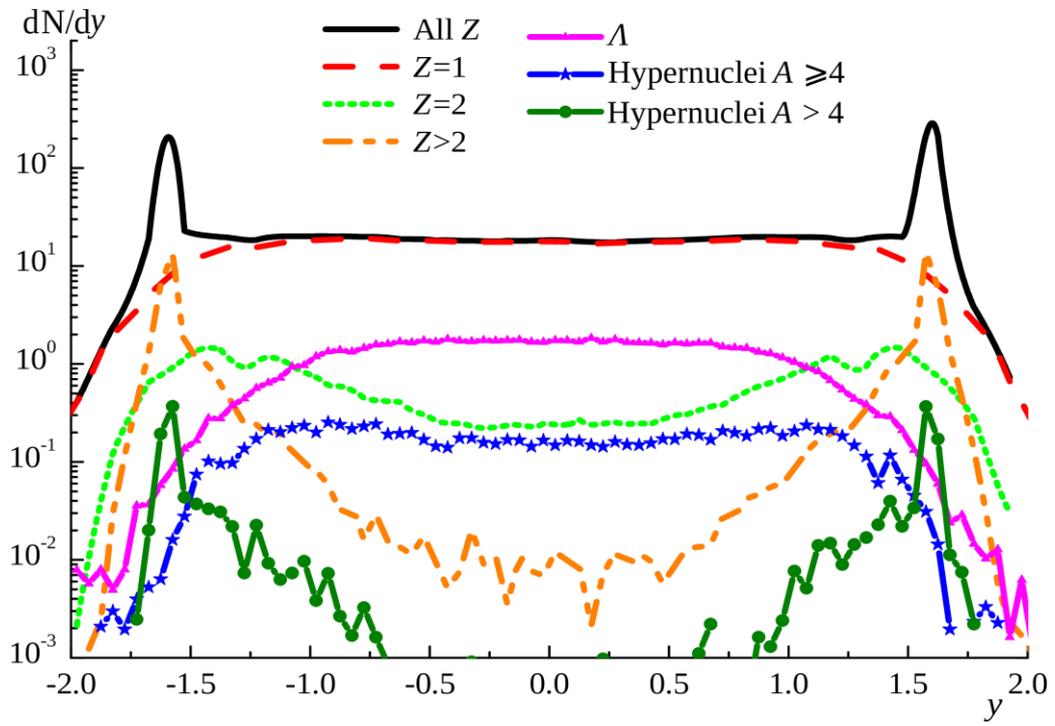